\documentclass[article]{aa}  
\usepackage{graphicx,upgreek}
\usepackage{txfonts,bm,xcolor}
\usepackage[export]{adjustbox}
\usepackage[colorlinks,citecolor=blue]{hyperref}
\usepackage{csvsimple}
\usepackage{subfigure}
\usepackage{animate}
\begin{document} 
\title{Coherent radio emission from a population of RS Canum Venaticorum systems}
\titlerunning{Coherent radio emission from RS\,CVns}
\authorrunning{Toet, S.~E.~B. et al.}
\author{S.~E.~B. Toet$^{1,2}$,
H.~K. Vedantham$^{2,3}$,
J.~R. Callingham$^{1,2}$,
K.~C. Veken$^{1,2}$,
T.~W. Shimwell$^{1,2}$,
P.~ Zarka$^{4,5}$,
H.~J.~A.~R\"{o}ttgering$^{1}$,
A. Drabent$^{6}$,
}

\institute{
$^1$Leiden Observatory, Leiden University, PO Box 9513, 2300 RA, Leiden, The Netherlands\\
$^2$ASTRON, the Netherlands Institute for Radio Astronomy, Oude Hoogeveensedijk 4,7991 PD Dwingeloo, The Netherlands\\
$^3$Kapteyn Astronomical Institute, University of Groningen, Postbus  800, 9700 AV Groningen, The Netherlands \\
$^4$Station de Radioastronomie de Nan\c{c}ay, Observatoire de Paris, PSL Research University, CNRS, Universit\'{e} Orl\`{e}ans, OSUC, 18330 Nan\c{c}ay, France\\
$^5$LESIA, Observatoire de Paris, CNRS, PSL, Meudon, France\\
$^6$Th{\"u}ringer Landessternwarte, Sternwarte 5, D-07778 Tautenburg, Germany\\
}

\date{}

 
  \abstract{
Coherent radio emission from stars can be used to constrain fundamental coronal plasma parameters, such as plasma density and magnetic field strength. It is also a probe of chromospheric and magnetospheric acceleration mechanisms. Close stellar binaries, such as RS~Canum~Venaticorum (RS\,CVn) systems, are particularly interesting as their heightened level of chromospheric activity and possible direct magnetic interaction make them a unique laboratory to study coronal and magnetospheric acceleration mechanisms.
RS\,CVn binaries are known to be radio-bright but coherent radio emission has only conclusively been detected previously in one system. Here, we present a population of $14$ coherent radio emitting RS\,CVn systems. We identified the population in the ongoing LOFAR Two Metre Sky Survey as circularly polarised sources at 144\,MHz that are astrometrically associated with known RS\,CVn binaries. We show that the observed emission is powered by the electron cyclotron maser instability. We use numerical calculations of the maser's beaming geometry to argue that the commonly invoked `loss-cone' maser cannot generate the necessary brightness temperature in some of our detections and that a more efficient instability, such as the shell or horseshoe maser, must be invoked. Such maser configurations are known to operate in planetary magnetospheres. We also outline two acceleration mechanisms that could produce coherent radio emission, one where the acceleration occurs in the chromosphere and one where the acceleration is due to an electrodynamic interaction between the stars. We propose radio and optical monitoring observations that can differentiate between these two mechanisms.
}


\maketitle
%
\section{Introduction}
Coherent radio emission from stellar systems can be produced by two mechanisms: plasma emission that occurs at the ambient plasma frequency and its harmonics, and cyclotron emission that occurs at the ambient cyclotron frequency and its harmonics \citep{MelroseandDulk1982}. Since plasma and cyclotron frequencies depend only on plasma density and magnetic field strength respectively, radio observations provide a unique opportunity to determine fundamental plasma parameters in the coronae of stars and magnetospheres of planets.

Both plasma and cyclotron emission are expected to be circularly polarised \citep{Dulk_review}, whereas the vast majority of radio sources are weakly ($\lesssim$1\%) or not circularly polarised and of extragalactic origin. We exploited the fact that the majority of radio sources are not circularly polarised to conduct an unbiased survey of stellar radio emissions via cross-matching circularly polarised radio sources in the ongoing LOFAR Two-Meter Sky Survey \citep[LoTSS;][]{LoTSS} with known stars \citep{Vedantham_Callingham_2020,Callingham2021}. Here we present the discovery and initial characterisation of 14 radio-emitting RS~Canum~Venaticorum (RS\,CVn) binaries detected by this technique. Our detections of individual main-sequence stars and wide-separation binaries will be presented elsewhere. With only a singular instance of coherent radio emission from an RS\,CVn binary known before \citep{HR_1099}, our sample represents a significant advance towards employing radio observations to study the dynamics of plasma in close stellar binaries.

RS\,CVn binaries usually consist of two stars with late spectral types (F,G or K) on the main sequence \citep{RS_CVns}. They are  short-period binaries, with periods ranging from $\sim$1 day to $\sim$30 days. The individual stars are in close proximity to each other ($\sim$0.01 AU). The systems also display strong Ca\,II H and K emission lines, and a prominent Balmer H$\alpha$ line. The prominent lines are due to an active chromosphere, which is why these stars are often called chromospherically active binaries. They are tidally locked, which leads to high rotational velocities, which in turn leads to strong magnetic fields \citep{Tidally_Locked,Reiners_2014}. These high velocities twist the magnetic flux loops in the stars, extending them through the photosphere at various points, leading to a large portion (up to 30\%) of the surface to be covered by starspots \citep{RS_CVn_starspots}. Such a large spot covering is used to explain the large optical photometric variability observed for these stars \citep{RS_CVn_new_class}.

Chromospheric flares, as well as magnetic interactions within the binary can accelerate electrons that emit radio waves \citep{Dulk_review,HR_1099}. While their magnetic interaction is interesting in its own right, it also provides a laboratory to study a more general class of sub-Alfv\'{e}nic plasma interactions that can occur between a star and its planet or a planet and its moon \citep{goldreich1969,zarka2007,lanza2009,saur2013}. 

Polarised and unpolarised radio emission from RS\,CVns has primarily been studied at gigahertz frequencies \citep{RS_CVn_typical_field_strength, II_Pegasi_circular_polarization,1995ApJ...444..342W, 1994A&A...286..181V, 1989ApJS...71..905D, 1988AJ.....95..204M, 1987AJ.....93.1220M, 1987ApJ...312..278W, 1985AJ.....90..493M}. The observed gigahertz emission is typically long-lasting and weakly polarised, and is interpreted as gyrosynchrotron radiation \citep{Dulk_review, 1985A&A...149..343K, 1990Ap&SS.170...81C}. Only in a few instances has long-lasting ($\gtrsim $hours) polarised ($\gtrsim$40\%) emission, reminiscent of planetary aurorae, been seen. For example, \citet{1995ApJ...444..342W} interpreted the different circular-polarised handedness between 1 and 8\,GHz observations as a suggestion that a coherent emission mechanism is likely operating at low frequencies. Furthermore, \citet{HR_1099} postulated that this emission might be the result of an electrodynamic binary interaction that accelerates electrons which cause the emission--- a scenario that shares similarities with the well-known Jupiter-Io `unipolar-induction' model \citep{goldreich1969}.

Radio observations at metre-wavelengths provide the best opportunity to test the electrodynamic interaction models for two reasons. Firstly, the surface cyclotron frequency $\nu_e \approx 2.8 \times 10^6 B$ Hz, above which the radio emission cuts off, in the vast majority of stars in RS\,CVn binaries, may lie below the gigahertz regime. This frequency limit is due to typical surface magnetic fields being around $B \approx 100$ Gauss \citep{RS_CVn_typical_field_strength}. In this case, only sub-gigahertz observations would be able to access their cyclotron emission from electrodynamic interaction. Secondly, cyclotron emission is inherently beamed, which means only a small fraction of radio-loud systems are visible to an observer. Low-frequency telescopes have wider fields of view than gigahertz-frequency telescopes, providing a high survey rate to identify the sources that have a fortuitous viewing geometry. The ongoing LoTSS survey \citep{LoTSS}, whose data we use here, is particularly well-suited for this purpose. Its long eight-hour exposures allow us to identify long-duration auroral-type emission against minute-timescale bursts that can occur on Sun-like stars due to flaring activity.

The rest of the paper is organised as follows. In \S2, we present the radio detections, fundamental properties of the RS\,CVns detected, and arguments in favour of a cyclotron maser interpretation. In \S3, we model the emission geometry and brightness temperature of our detections using theoretical considerations and numerical calculations of the emitter's beaming geometry. We also outline two plausible mechanisms that accelerate charges that radiate the observed radio waves. We end in \S4 with a summary and outlook towards determining the mechanism of acceleration in RS\,CVn systems.

\section{Radio observations and properties}
\subsection{Discovery}

LoTSS routinely produces 144\,MHz images with $\lesssim$100$\,\upmu$Jy root-mean-square noise and with an astrometric accuracy of $\sim$0$\arcsec.2$ \citep{LoTSS}, making it an unparalleled widefield low-frequency survey to blindly search for radio-bright stellar systems. To discover coherently-emitting stellar systems, and to dramatically minimise our false-positive association rate \citep{Gaia_LoTSS_Association}, we isolated our crossmatch to only circularly-polarised sources in LoTSS. The number density of sources with a circularly-polarised fraction $>2\%$ is $\approx0.1$ per sq.~degree \citep{Callingham2021}. After identifying $\geq 4\sigma$ circularly-polarised LoTSS sources, we associated a \emph{Gaia} data release 2 (DR2) \citep{Gaia_DR2} counterpart if the \emph{Gaia} DR2 source was within 1$''$ of the radio position and had a parallax over error of $\varpi / \varpi_{\mathrm{err}} \geqslant 3$. Since the LoTSS data available at the time of writing only covers $\approx$20\% of the northern sky at Galactic latitudes $|b| > 20^{\circ}$, where the \emph{Gaia} DR2 source density is relatively low, the chance-coincidence associations between a Gaia Galactic source and the Stokes\,V LoTSS sample is $\approx0.05$ \citep{Callingham2021}. Therefore, we consider our associations highly reliable. 

\begin{figure*}
\includegraphics[width=\linewidth]{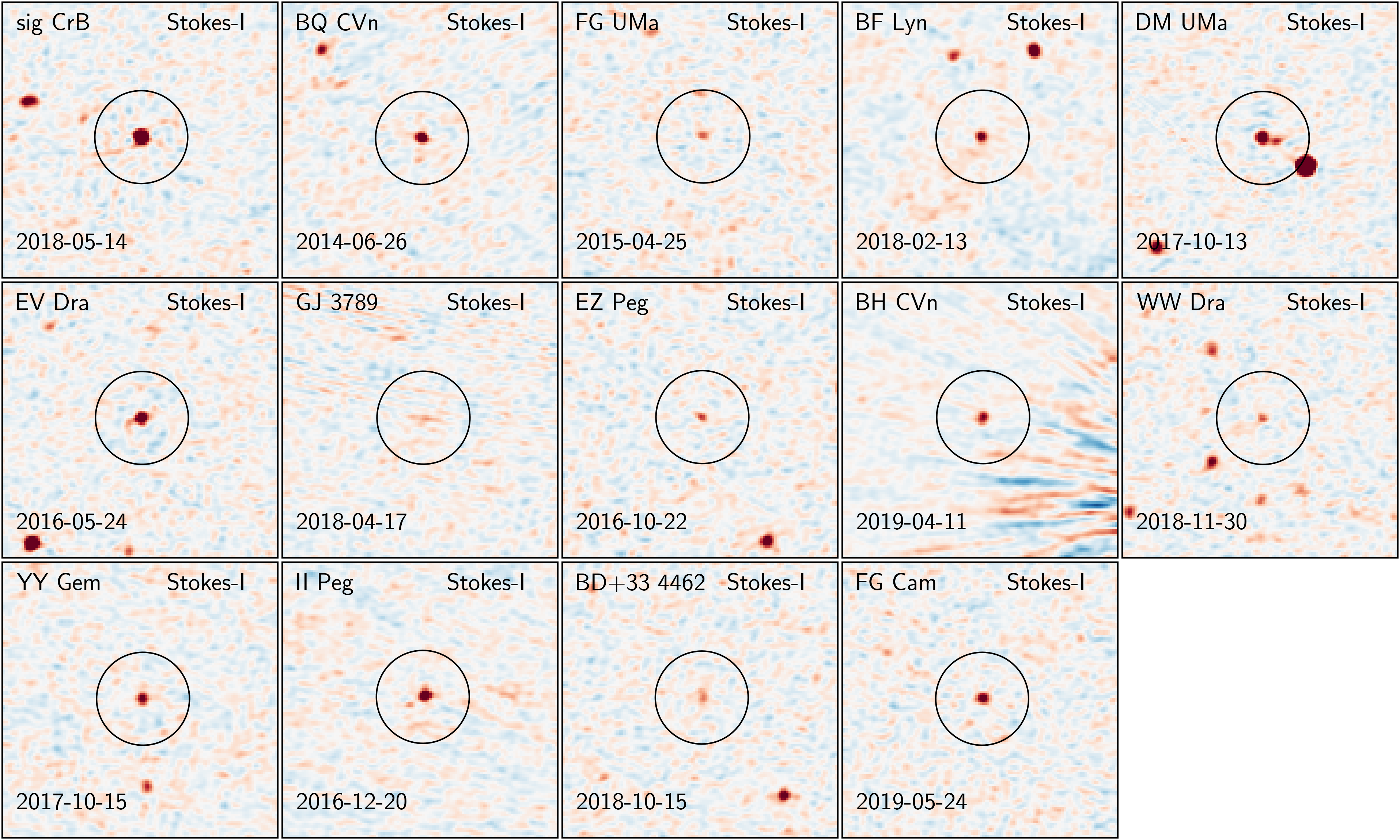}
\caption{Stokes I discovery images of our RS\,CVn population. Each panel is a zoom-in of a LoTSS pipeline-produced image of the field containing the target. The circles are centred on the proper-motion corrected position of the targets and have a radius of $30''$. The point spread function is about $5''$ wide. The color scale runs from $-15$ (blue), through $0$ (white), to $+15$ (red) median absolute deviations, which corresponds to a range of $-10\sigma$ to $+10\sigma$ for Gaussian noise. The date reported in each panel is the date of observation in YYYY-MM-DD format.\label{fig:I}}
\end{figure*}

\begin{figure*}
\includegraphics[width=\linewidth]{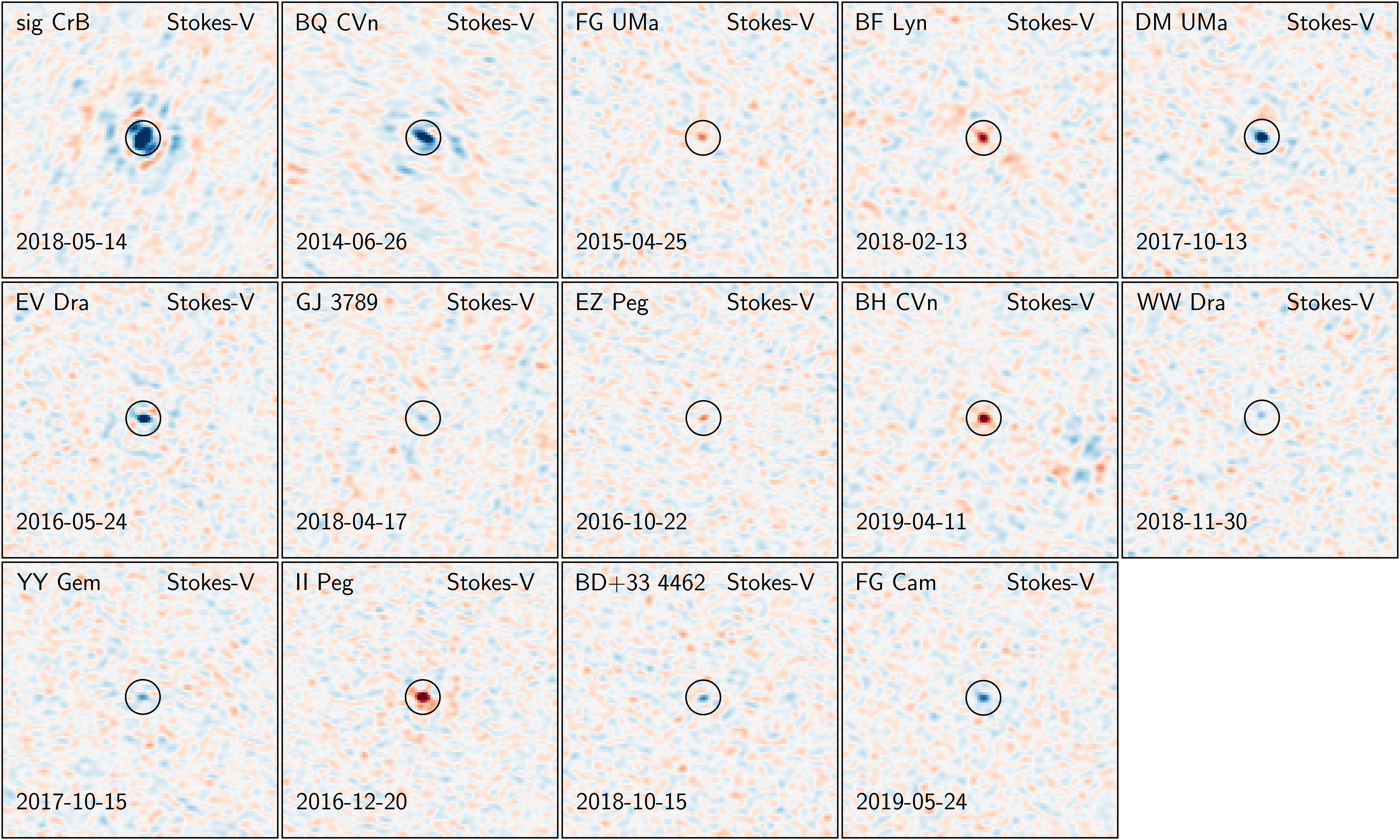}
\caption{Same as Fig. \ref{fig:I}, but for Stokes V. The point spread function width is about $20''$ and the radius of the circle is $30''$. The images are not de-convolved.\label{fig:V}}
\end{figure*}

From this sample of coherently-emitting stellar systems with \emph{Gaia} DR2 counterparts, we identified three classes of objects: isolated (or wide binary) M\,dwarfs \citep[presented in ][]{Callingham2021}, millisecond pulsars, and different types of variable stars such as the FK Comae Berenices (FK\,Com), W Ursae Majoris (W\,UMa), BY Draconis (BY\,Dra), and RS\,Canum Venaticorum (RS\,CVn) types. Since the main science goal of this manuscript is associated with binary interactions, we only present the stellar systems that are known to be close binaries with periods less than tens of days. This list includes binaries classified as RS\,CVn and BY\,Dra systems. We plan to present the other sources in a followup paper. The LoTSS photometry of our sample is given in Table\,\ref{Table_Observables}, while Figures \ref{fig:I} and \ref{fig:V} show the Stokes I and Stokes V discovery images. The known stellar parameters and orbital parameters of the individual systems are given in Table \ref{Table_Stellar_Parameters} and Table\,\ref{Table_Orbital_Parameters} respectively.


\subsection{Images and lightcurves}

Once a source of interest was identified in the LoTSS data, we performed some post-processing to extract the highest signal-to-noise images and lightcurves. The post-processing steps are described in \citet{2020arXiv201102387V} and were also employed in \citet{cr_dra} and \citet{ Vedantham_Callingham_2020}. Briefly, each dataset was phase shifted to the position of the source of interest, taking account of the LOFAR station beam attenuation. Residual phase errors were solved for by applying a self-calibration loop on 10-20\,s timescales. This was followed by several rounds of  diagonal gain calibration on timescales of $\approx$20\,min. 

Stokes I and V images of each source were then created from these reduced datasets using a robust parameter of $-$0.5 via \textsc{WSClean} \citep[v\,2.6.3;][]{2014MNRAS.444..606O}. For the time series analysis, each image was synthesised over all of the available bandwidth to maximise the signal. The flux density of each source was measured using the Background And Noise Estimator (\textsc{BANE}) and source finder \textsc{Aegean} \citep[v\,2.1.1;][]{2012MNRAS.422.1812H,2018PASA...35...11H}. We used the priorised fitting option of \textsc{Aegean} at the location of each source to account for uncertainties associated with non-detections in our time series analysis.

Our time sampling function of the radio data for each stellar system was determined by the Stokes I detection in each 8 hour integrated epoch and the period of the system. The lightcurves of some of our brightest detections and shortest-period binaries are discussed in more detail in Section\,\ref{sec:lightcurves}.

\subsection{Mechanism of emission}
We determined the mechanism of emission using the approximate brightness temperature and degree of circular polarisation. 
The brightness temperature, normalised to characteristic values for stars in our sample is 
\begin{equation}
\label{eqn:tbobs}
    T_b \approx 2\times 10^{12} \left(\frac{d}{50\,{\rm pc}}\right)^2 \left(\frac{R}{R_\odot}\right)^{-2}\left( \frac{S}{1\,{\rm mJy}}  \right)\,\,{\rm K},
\end{equation}
where $d$ is the distance to the star, $S$ is the radio flux-density, and $R$ is the projected radius of the emitter (assumed to be circular). Conservatively, using the values of $R$ for the larger binary component, we find that $T_b$ ranges from $\sim 10^{11}\,{\rm K}$ to $\sim 10^{12.5}\,{\rm K}$ in our sample.
The high implied brightness temperature, in combination with the high circular polarisation fraction, implies the radio emission can not be produced by any incoherent emission mechanism \citep{Dulk_review}. Therefore, only coherent plasma and cyclotron maser emission are plausible emission mechanisms. 

Plasma emission is thought to be caused by conversion of Langmuir waves to transverse electromagnetic waves \citep{ginzburg1958,melrose1991}. At low radio frequencies ($\nu\sim 10^2\,{\rm MHz}$), induced emission can be neglected for reasonable coronal parameters \citep{Vedantham2021} and we arrive at the following expressions for the anticipated brightness temperature of fundamental and second harmonic plasma emission, respectively \citep{Vedantham2021}
\begin{equation}
    T_{b,f}\sim 2 \times 10^{10} \left(\frac{\nu}{10^8\,{\rm Hz}}\right)\left(\frac{h_p}{10^{10}}\right)\left(\frac{T_1}{10^8}\right)\left(\frac{w}{10^{-5}}\right) \,{\rm K},
    \label{Plasma_Fundamental}
\end{equation}
\begin{align}
    T_{b,h}\sim 4 \times 10^{13} \left(\frac{h_p}{10^{10}}\right) \left(\frac{T}{10^{6.5}}\right)^4\left(\frac{T_1}{10^8}\right)^{-1/2}\left(\frac{w}{10^{-5}}\right)^2 \,{\rm K}.
    \label{Plasma_Second_Harmonic}
\end{align}
Here $h_p$ is the plasma density scale height, $T_1$ is the hot plasma component temperature that drives the generation of Langmuir turbulence via the bump-on-tail instability, $T$ is the ambient plasma temperature, and $w$ is the fractional energy in the Langmuir-wave turbulence. The normalisation values are chosen to be representative values. The turbulence level, $w$ is thought to be limited to $\sim 10^{-5}$ at which point the Langmuir wave packets collapse due to non-linear effects \citep{benzbook}. The ambient coronal temperature of RS\,CVns determined from X-ray observations is $T\approx 10^{6.5}$K \citep{ness2003} which gives a density scale height of $h_p\approx 0.3R_\ast$. $T_1$ has not been observationally determined for our targets and we have taken $T_1=30T$ as a typical value based on hard X-ray observations of flares on dMe stars and RS\,CVn binaries \citep{2001A&A...375..196F,1999A&A...350..900F}.

Although second harmonic plasma emission can attain the observed brightness temperature (see equations \ref{eqn:tbobs} and \ref{Plasma_Second_Harmonic}), it cannot attain the high levels of observed circular polarisation \citep{Vedantham2021} and is therefore disfavoured. Fundamental plasma emission can, in principle, be 100\% circularly polarised, but it cannot attain the observed brightness temperature unless $R\gtrsim$ a few to ten $R_*$ for stars in our sample  (see equations \ref{eqn:tbobs} and \ref{Plasma_Fundamental}). Because we are detecting quasi-quiescent emission in eight-hour-long exposures, the source size must be smaller than the closed-field region in the corona where turbulent gas can be trapped by the magnetic field. In other words, the source must lie within the Alfv\'{e}n surface. We can approximately determine the Alfv\'{e}n radius by equating the plasma and magnetic energy densities. For a characteristic base plasma density of $n_0=10^{10}\,{\rm cm}^{-3}$ \citep{ness2003} and a dipolar magnetic field with a surface field strength of $B$, the Alfv\'{e}n point is at at a radial distance of $R_{\rm A} \approx 2R_\ast (B/100\,{\rm G})^{1/3}$. Hence confining turbulent plasma at a radial distance of $\approx 10R_\ast$ is theoretically problematic which disfavours the plasma emission hypothesis for a significant fraction of our sample. Although this leads us to favour the cyclotron maser as the mechanism of emission, the plasma hypothesis cannot be conclusively ruled out because our calculations are based only on nominal values for parameters such as $w$, $T$, $T_1$ and $h_p$.
We note here that \citet{HR_1099} also identify the electron cyclotron maser as the mechanism of emission for their observations of HR\,1099 at gigahertz frequencies. While they arrived at this conclusion based on the detection of millisecond-timescale structure in the radio emission, the low frequency of our observations has allowed us to use the brightness temperature and polarisation fraction of the hours-long-duration emission to arrive at the same conclusion. Although \citet{HR_1099} only had a sample of one, our larger RS\,CVn sample size suggests that close stellar binaries can generate quasi-quiescent cyclotron maser emission. This conclusion can be reconciled with previous attribution of gigahertz frequency observations mainly to gyrosynchrotron emission, if the large-scale magnetic fields of RS\,CVn stars have surface values well below a kiloGauss. In this case, their cyclotron emission does not extend to high enough frequencies to have been observed before. 

\begin{table*}
    \centering
    \csvreader[head to column names, before reading=\begin{adjustbox}{max width=\linewidth},
    after reading=\end{adjustbox},
    tabular={|l|l|l|l|l|},
    table head=\hline {\bfseries Star Name} & {\bfseries \bm{$S_I$} (mJy)} & {\bfseries \bm{$S_V$} (mJy)} & {\bfseries CPF (\%)} & {\bfseries Observation Start} \\\hline,late after line=\\\hline]{Observables_mJy.csv}{}
    {\StarName & \StokesI & \StokesV & \CPF & \ObservationStart}
    \caption{144\,MHz flux densities (Stokes I and V) of our population. The flux uncertainties are typically $0.1\,{\rm mJy}$. The Stokes V sign convention is similar to that used in Pulsar observations \citep{vanstraten} and opposite of the IAU convention \citep{hamaker1996}. Each observation lasted eight hours.}
    \label{Table_Observables}
\end{table*}

\setcitestyle{numbers}

\begin{table*}
    \centering
    \csvreader[head to column names, before reading=\begin{adjustbox}{max width=\linewidth},
    after reading=\end{adjustbox},
    tabular={|l|l|l|l|l|l|l|l|l|},
    table head=\hline {\bfseries Star Name} & {\bfseries Star Type} & {\bfseries Distance (pc) \citep{Gaia_DR2}} & {\bfseries Spectral Type} & {\bm{$R_{HC} \: (R_\odot)$}} & {\bm{$R_{CC} \: (R_\odot)$}} & {\bm {$M_{HC} \: (M_\odot)$}} & {\bm {$M_{CC} \: (M_\odot)$}} \\\hline,late after line=\\\hline]{Stellardata.csv}{}
    {\StarName & \StarType & \Distance & \SpectralType & \RHC & \RCC & \MHC & \MCC}
    \setcitestyle{authoryear}
    \caption{This Table shows relevant stellar properties found in the literature search for the stars detected by LOFAR. The stars are in order of processing date. $R_{HC}$ is the radius of the hot component (earlier spectral type), $R_{CC}$ is the radius of the cold component (later spectral type). $M_{HC}$ is the mass of the hot component and $M_{CC}$ is the mass of the cold component. Radii denoted with an asterisk ($\text{*}$) are estimated by taking the known radius of a star with the same spectral type \citep{Stellar_Diameters_II,Stellar_Diameters_III,FI_Cnc_Spectral_Type}. The masses of BD+334462 are $M\sin^3(i)$ due to unknown inclination, and are therefore lower limits. All distance measures come from Gaia DR2, except for Sig CrB which is only available in the Hipparcos catalogue and FG Cam which is only available in Gaia DR1. DM UMa is not spectrally resolved, therefore we give the combined spectral type. The literature search was made significantly more convenient thanks to the catalogue of Chromospherically Active Binaries \citep{CABcatalogue}. The references corresponding to the numbers are given in Table \ref{Reference_table}.}
    \label{Table_Stellar_Parameters}
\end{table*}

\begin{table*}
    \centering
    \csvreader[head to column names, before reading=\begin{adjustbox}{max width=\linewidth},
    after reading=\end{adjustbox},
    tabular={|l|l|l|l|l|l|l|},
    table head=\hline {\bfseries Star Name} & {\bm{$P_{rot}$} \bfseries(days)} & {\bm{$P_{orb}$} \bfseries(days)} & {\bm{$a \: (R_\odot)$}} & {\bm{$a_{Kep} \: (R_\odot)$}} & {\bfseries Inclination (deg)} & { \bfseries Eccentricity} \\\hline,late after line=\\\hline]{Orbital_Parameters.csv}{}
    {\StarName & \Prot & \Porb & \SMA & \SMAKep & \Inc & \Ecc}
    \setcitestyle{authoryear}
    \caption{The orbital parameters of our targets. $P_{rot}$ is the rotational period; The values displayed here are literature photometric periods, but we assume them to be equal to the rotational periods of the individual stars \citep{FK_Com_Spectral_Type_Starspots}, which are assumed to be equal to one another due to tidal locking \citep{Tidally_Locked}. $P_{orb}$ is the orbital period, $a$ is the semi-major axis from the literature and $a_{Kep}$ is the semi-major axis computed using the masses in Table \ref{Table_Stellar_Parameters}, the orbital period and Kepler's Third Law: $P_{orb}^2/a_{Kep}^3=4\pi^2/G(M_{HC}+M_{CC}$). For this we of course needed both masses and the orbital period. If one of these was unknown, we indicated this with `-'. The Keplerian semi-major axis is then also unknown and indicated with '-'. Only $a\sin(i)$ and $Msin^3(i)$ were known for BD+334462, explaining the lower limit on $a$ and on $a_{Kep}$. The references corresponding to the numbers are given in Table \ref{Reference_table}.}
    \label{Table_Orbital_Parameters}
\end{table*}

\setcitestyle{authoryear}

\section{Emission modelling}
\label{Discussion}
\subsection{Energetics}
The isotropic-equivalent radio luminosity of our sample is of the order $\mathcal{E}\sim 10^{24}\,{\rm ergs/s}$, which is many orders of magnitude below the typical X-ray luminosity of RS\,CVn coronae. Therefore, chromospheric flares, that are thought to power the X-ray corona, certainly have sufficient energy to power the radio emission. Next, we consider models that posit the magnetic interaction between the stars as the source of electron acceleration. RS\,CVns are likely tidally locked \citep{Tidally_Locked}, but deviate from strict magnetospheric co-rotation by about $\eta = 1\%$ \citep{hall1981}. This differential motion can potentially accelerate electrons and power the emission. Let $\Omega$ be the rotation rate, $a$ the orbital separation, $B_0$ the surface magnetic field of the emitter, and $R_1$ its radius. The field at the conducting obstacle (binary companion) is $B = B_0(a/R_1)^{-3}$. The magnetic energy intercepted by the companion of radius, say $R_2$, per unit time is $\eta a\Omega B^2/(8\pi)\pi R_2^2$ which in convenient units for a Keplerian relationship between $a$ and $\Omega$, and assuming solar mass stars is
\begin{equation}
\label{eqn:Emag}
    \mathcal{E} \sim 10^{24} \left(\frac{\eta}{0.01}\right) \left(\frac{R_1}{R_\odot}\right)^6\left(\frac{R_2}{R_\odot}\right)^2\left(\frac{B_0}{100\,{\rm G}}\right)^2 \left(\frac{a}{10R_\odot}\right)^{-6.5}\,{\rm ergs/s.}
\end{equation}

Alternatively, the intercepted energy flux may be dominated by the ram-pressure of the stellar wind fluid instead of the energy stored in the magnetic field. In this case we have $\mathcal{E} \approx \pi m_p n v^3 R_2^2$. where $m_p$ is the proton mass, $n$ is the stellar wind density at the location of the companion, and $v$ is the wind velocity. We make the simplifying assumption that the wind has reached its terminal velocity which gives $n = n_0 (a/R_1)^{-2}$ where $n_0$ is the base coronal plasma density. In convenient units, we get
\begin{equation}
\label{eqn:Eram}
    \mathcal{E} \sim 10^{25}\left(\frac{n_0}{10^{10}\,{\rm cm}^{-3}}\right) \left(\frac{R_1R_2}{R_\odot^2}\right)^{2}\left(\frac{a}{10\,{\rm AU}}\right)^{-2}\left(\frac{v}{600\,{\rm km/s}}\right)^{3}\,{\rm ergs/s}.
\end{equation}

We shall now compare these estimates to the power necessary to support the radio emission. 
Assuming the radio emission to have a beam solid angle of $10^{-1}$\,sr, comparable to that of Jovian emission due to Io's interaction \citep{queinnec1998} and an emission bandwidth going up to the surface cyclotron cut-off of $2.8(B_0/{\rm G})$\,MHz the inferred radio power is
\begin{equation}
\label{eqn:Ereq}
    \mathcal{E} \sim 10^{22} \left(\frac{d}{100\,{\rm pc}}\right)^2\left(\frac{S}{1\,{\rm mJy}}\right)\left(\frac{B_0}{100\,{\rm G}}\right)\,{\rm ergs/s}.
\end{equation}
A quick comparison of equations \ref{eqn:Emag}, \ref{eqn:Eram} and \ref{eqn:Ereq} with the tabulated properties of our RS\,CVns systems shows that there is sufficient intercepted magnetic pressure and ram pressure in the binary interaction to drive the radio emission if the radio emission efficiency is $\gtrsim  10^{-2}$ and $\gtrsim 10^{-3}$ respectively. A comparison of the radio power of solar system planets and the interplanetary magnetic pressure as well as the solar wind's ram pressure yields efficiencies of $2\times 10^{-3}$ and $10^{-5}$ respectively \citep{zarka2007,zarka2001}. Therefore, although both mechanisms are energetically feasible in principle, the empirical scaling law from solar system planets suggests that  magnetic interaction is causing the radio emission. We note here that the X-ray luminosity of RS\,CVns is typically $\gtrsim 10^{32}\,{\rm ergs/s}$ which is clearly not powered by such interactions; a point also noted by \citet{schrijver1991}. Indeed, the cyclotron emission comes from a small population of supra-thermal ($\sim$10 keV to $\sim$1 Mev) electrons that represent a minuscule fraction of the overall coronal thermal-energy budget. 

\subsection{Brightness temperature}
We next check if the cyclotron maser mechanism can plausibly generate the necessary brightness temperature by comparing theoretical limits with observationally inferred values for our sample. Because cyclotron emission is inherently beamed (emitted nearly perpendicular to the magnetic field) and narrowband, the `observed' brightness temperature must itself be inferred via computation of the projected source size, $A$, at any given frequency. 

We computed the source size taking beaming into consideration numerically (details in the Appendix), but here provide an approximate analytic expression to develop intuition.
For simplicity, let us consider a large scale dipolar structure for the magnetosphere of the emitter. 
Cyclotron maser emission from the loss-cone instability is beamed into the surface of a cone with a large opening angle $\theta\lesssim \pi/2$ and narrow thickness $\Delta\theta\approx \beta$, where $\beta c$ is the speed of the emitting electrons \citep{MelroseandDulk1982}.  The cone-axis is parallel to the ambient magnetic field. The emission therefore fills a solid-angle fraction $\beta\approx \Delta\theta$ of the visible hemisphere. Seen conversely, a given observer can only view emission from the same fraction of the visible stellar disk. Hence, the apparent emitter area for brightness temperature calculation is $A = \pi\beta R_\ast^2$, which we confirm to be a good approximation using numerical calculations (see Appendix). Taking $\beta=0.2$ as a characteristic value \citep[based on Io-induced Jovian emission models ][]{hess2010}, we see that the observed brightness temperature from equation \ref{eqn:tbobs} is of the order $\sim 10^{12}-10^{13.5}\,{\rm K}$, the upper end of which is comparable to the theoretical limit for maser emission driven by the loss-cone instability \citep{MelroseandDulk1982}. In other words, a loss-cone maser requires the instability to operate in a large fraction (of order unity) of a dipolar magnetosphere. However, the emitting region in known instances of the loss-cone maser, such as planetary auroral emission, is confined to a thin oval around the magnetic poles. Furthermore, if the loss-cone instability were omnipresent it should have made most of our targets continuously visible because at any given instant, there would be a part of the magnetosphere whose magnetic field orientation allows for the emission to be beamed towards the Earth. 

\begin{figure*}
\centering
\includegraphics[width=\linewidth]{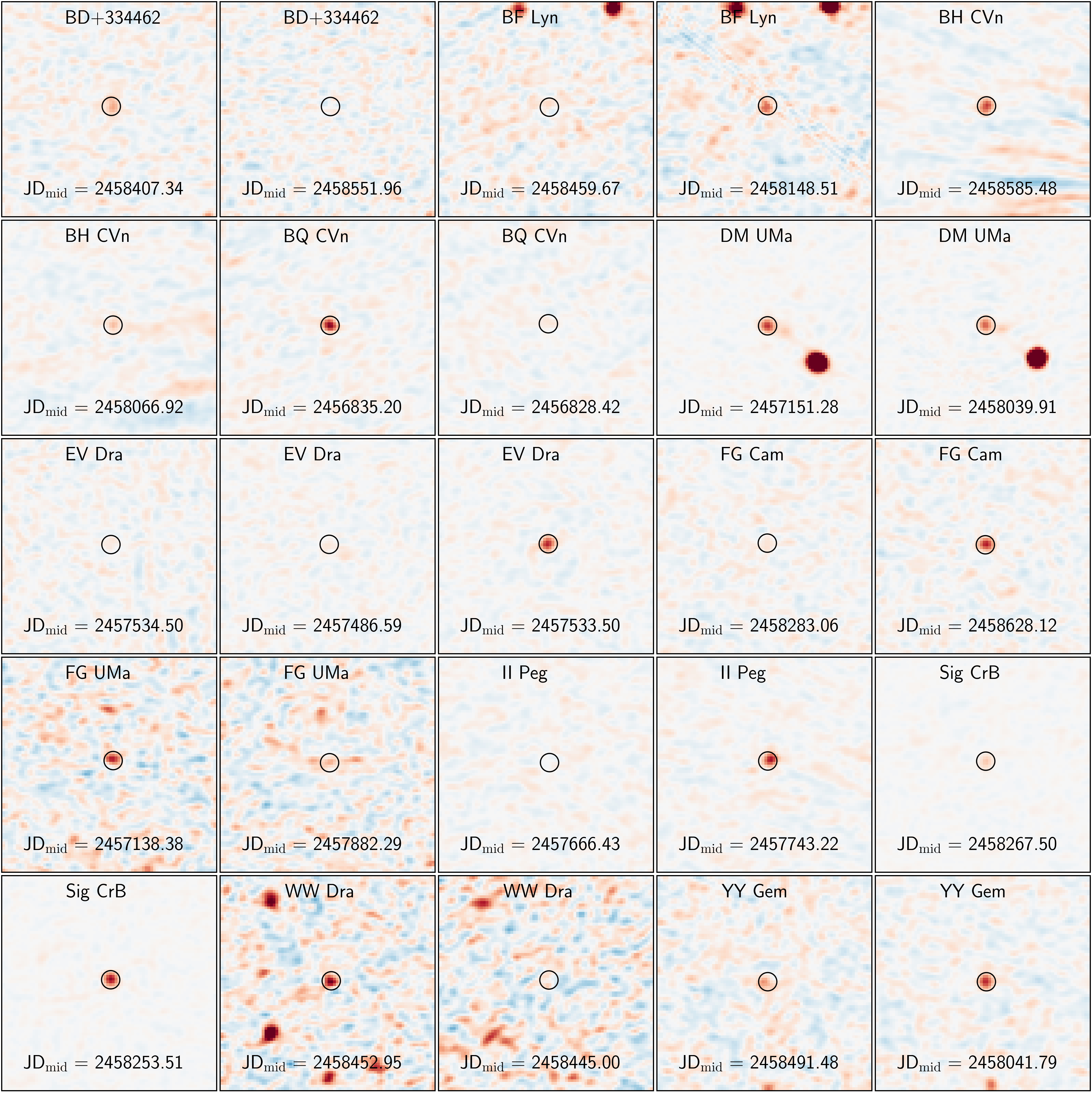}
\caption{Multi-epoch Stokes-I images of our targets. The color scale in each panels runs from $-S$ (blue) through zero (white) to $+S$ (red) where $S$ is the flux density in the discovery epoch (see Table \ref{Table_Observables}). The annotated text also provides the Julian date at the mid-point of the eight-hour exposure. The circular target-marker has a radius of $5''$. The full-width at half-maximum of the point spread function is about $5''$. EZ Peg and DG CVn are not included here because both only had one detection of sufficient sensitivity.
\label{fig:multi-epoch}}
\end{figure*}

To test if this is the case, we exploited the fact that some of our targets were observed by multiple neighbouring LoTSS pointings. Fig. \ref{fig:multi-epoch} shows a montage of images of our targets where such multi-epoch data with sufficient signal-to-noise ratio for detection was available. We find that most of our targets are detected only in a single epoch which suggests that the maser sites do not occupy a large fraction of the magnetosphere in these objects. We therefore disfavour the loss-cone instability as the cause of emission, particularly for systems at the upper end of the observed brightness temperature range in our sample. 

Let us now consider models where only a small fraction of the magnetosphere of the emitting star is `live' with cyclotron emitting electrons. In the specific model of Jupiter-Io type interaction,  only the magnetic flux tube connecting the two bodies is live with emitting electrons (see Fig. \ref{Binary_Interaction_Gifs_Magnetic_Field_Lines} for an illustration). Because the loss-cone driven maser in such instances is infeasible due to the above brightness temperature arguments, we suggest that a more efficient instability such as the horseshoe or shell-type maser \citep{Horseshoe_Maser} is responsible. The source size in this case is best computed numerically (as shown in the Appendix). The beaming geometry of the horseshoe and shell-type masers is similar to that of the loss-cone maser (all emit nearly perpendicular to the magnetic field) \footnote{The level of refraction experienced by the radiation will vary between the masers, however this is beyond the scope of this paper.}. We therefore use the loss-cone maser code for our predictions in section \S3.4. Here we first provide an approximate expression for analytic intuition. Along a given live field-line, the source size at some frequency is limited by the evolution of the cyclotron frequency beyond the maser bandwidth. The scale length for magnetic field variation in a dipole is $\sim R_\ast/3$ and if the maser bandwidth is $\Delta\nu/\nu$, then the source size along the field lines is $(\Delta\nu/\nu)R_\ast/3$. Across the field lines, the source size is dictated by the thickness of the flux tube that has emitting electrons. Let the radial extent of this `live' flux tube in the equatorial plane be $\Delta L$. The thickness close to the emitting star's surface is $\sim \Delta L (L/R_\ast)^{-3/2}$. An approximate value for the source area at a given frequency is then
\begin{equation}
    A \sim R^2_\ast\frac{1}{3}\frac{\Delta \nu}{\nu} \frac{\Delta L}{R_\ast} \left(\frac{L}{R_\ast}\right)^{-3/2}.
\end{equation}
For characteristic values of $\Delta\nu/\nu\sim 0.01$, $L/R_\ast\sim 10$, $\Delta L\approx R_\ast$ (size of the companion), we have $A/R_\ast^2\sim10^{-4}$. Hence, using Equation \ref{eqn:tbobs}, we obtain a maser brightness temperature of $T_b\sim 10^{15}-10^{17}\,{\rm K}$. A maser driven by the shell or horseshoe instability, thought to power the Earth's auroral kilometric radiation, can reach brightness temperatures of $\sim 10^{20}\,{\rm K}$ \citep{louarn1990,Ring_Shell_Brightness_Temperature} which can readily accommodate the inferred values.

In summary, we find that cyclotron maser emission from the loss-cone instability is unlikely to generate the necessary brightness temperature in some of our targets. Instead, the horseshoe of shell-type maser must be invoked to account for the estimated brightness temperature. We note, however, that the horseshoe and shell instabilities require a potential drop directed parallel to the magnetic field lines. It is presently unclear theoretically whether a sufficiently large field-aligned potential can be sustained in the corona of a star whose dense plasma will tend to `short' such voltage drops. In a planetary magnetosphere, the field-aligned potentials are created in density cavities. Although the same has been suggested for stellar coronae in the context of short bursts \citep{melrose2016}, whether such cavities can be long-lived to generate quasi-quiescent emission must await deeper theoretical work.

\subsection{Mechanism of acceleration}
\label{subsec:mechanism}
Although our blind survey detected a population of M-dwarfs \citep{Callingham2021}, we have not yet detected a significant population of isolated stars of earlier spectral types. The large number of RS\,CVn detections whose components span F, G and K stars clearly show that interaction with an orbital companion is essential to the observed radio emission.  We foresee two mechanisms by which binarity is causally related to the emission of detectable coherent radio emissions. (i) {\em Chromospheric acceleration:} The radio emission may simply be the result of significantly enhanced magnetic activity which in turn is caused by faster rotation (due to tidal locking) and the influence of tidal forces on the underlying dynamo (sometimes called breakdown of co-rotation) \citep{schrijver1991}. (ii) {\em Magnetospheric acceleration: } Alternatively, charge acceleration may be a direct result of electrodynamic interaction in the binary similar to the Jupiter-Io engine. In particular, the velocity shear between the wind of one star and the magnetosphere of the other can accelerate charges \citep{HR_1099}. In the first case, the acceleration site is located in the chromosphere and the mechanism is likely similar to that seen in isolated stars, whereas in the second case, it is located away from the star (i.e. it is external to the emitting star) and is likely similar to that seen in magnetospheres of gas-giant planets and brown dwarfs. 
Discriminating between these two mechanisms will be a significant step  in identifying the source of high-energy electrons in RS\,CVn systems. 

We propose that the geometric modulation of the observed emission is an avenue to discriminate between the two mechanisms stated above. Magnetospheric acceleration due to electrodynamic interaction will isolate the `live' azimuthal sectors to those containing the flux tube connecting the two stars, whereas chromospheric acceleration need not have the same preferred azimuthal sector. Due to the beamed nature of cyclotron maser emission, the chromospheric acceleration scenario will likely lead to emission that is always `on', whereas magnetospheric acceleration will lead to emission that has a periodic `on-off' modulation as the `live' flux tube's orientation changes with respect to the observer during the binary orbit.

There is an important caveat to the discriminant proposed above. Long term photometry of RS\,CVn starspots, that are the likely sites of particle acceleration, has shown that the spots tend to cluster in so-called Active Longitude Bands \citep[ALBs; ][]{alb}. Because the rotation and orbital periods of RS\,CVns are nearly identical, periodic modulation from this ALB-clustering could masquerade as magnetospheric acceleration. However, the starspot-dip in lightcurves of RS\,CVns where both stars are on the main sequence appear at quadrature points in the orbit (as opposed to conjunctions); the implication being that the active regions are separated in longitude from the line joining the two stars by $\pm 90^\circ$ \citep{olah2006}\footnote{In any case, starspot longitude can be measured with quasi-contemporaneous optical photometry}. On the other hand, the magnetospheric acceleration scenario posits that the longitude of the `live' flux tube is the same as that of the line joining the two stars. Therefore, even in presence of ALBs, canonical radio and starspot monitoring should be able to discriminate between the two scenarios for systems where both stars are on the main sequence.

\subsection{Application to short-period main-sequence binaries}
\label{sec:lightcurves}
Although we currently do not have long-term radio monitoring data, we applied the discriminant proposed in \S\ref{subsec:mechanism} to existing LoTSS data to elucidate the methodology. We chose a sub-set of our population with the following properties: (a) short orbital periods of a few days or less (semi-major axis of $\lesssim 10R_\odot$) for which the interaction is plausibly sub-Alfv\'{e}nic, (b) both stars on the main-sequence which avoids the ALB-phenomenon from masquerading as magnetospheric acceleration (see \S\ref{subsec:mechanism}), (c) systems for which the orbital inclination is known such that their cyclotron-maser visibility can be computed using known theory, and (d) systems for which the orbital ephemeris is known with sufficient accuracy so that the orbital phase at the radio observation epoch can be calculated with sufficient accuracy. With these criteria, we selected Sig\,CrB, BF\,Lyn, EV\,Dra and YY\,Gem to apply our discriminant between the two acceleration mechanisms. 

The orbital phase of the low-frequency radio emission, including epochs in which the source was not detected, and our model prediction for these four systems is presented in Figure\,\ref{fig:phasedig}. The predictions for the ten systems for which we had all necessary parameters are shown in Figure \ref{Emitter_Area_vs_Orbital_Phase}.

\subsubsection{Sig\,CrB}
\begin{figure*}
\begin{center}$
\begin{array}{cc}
\includegraphics[scale=0.24]{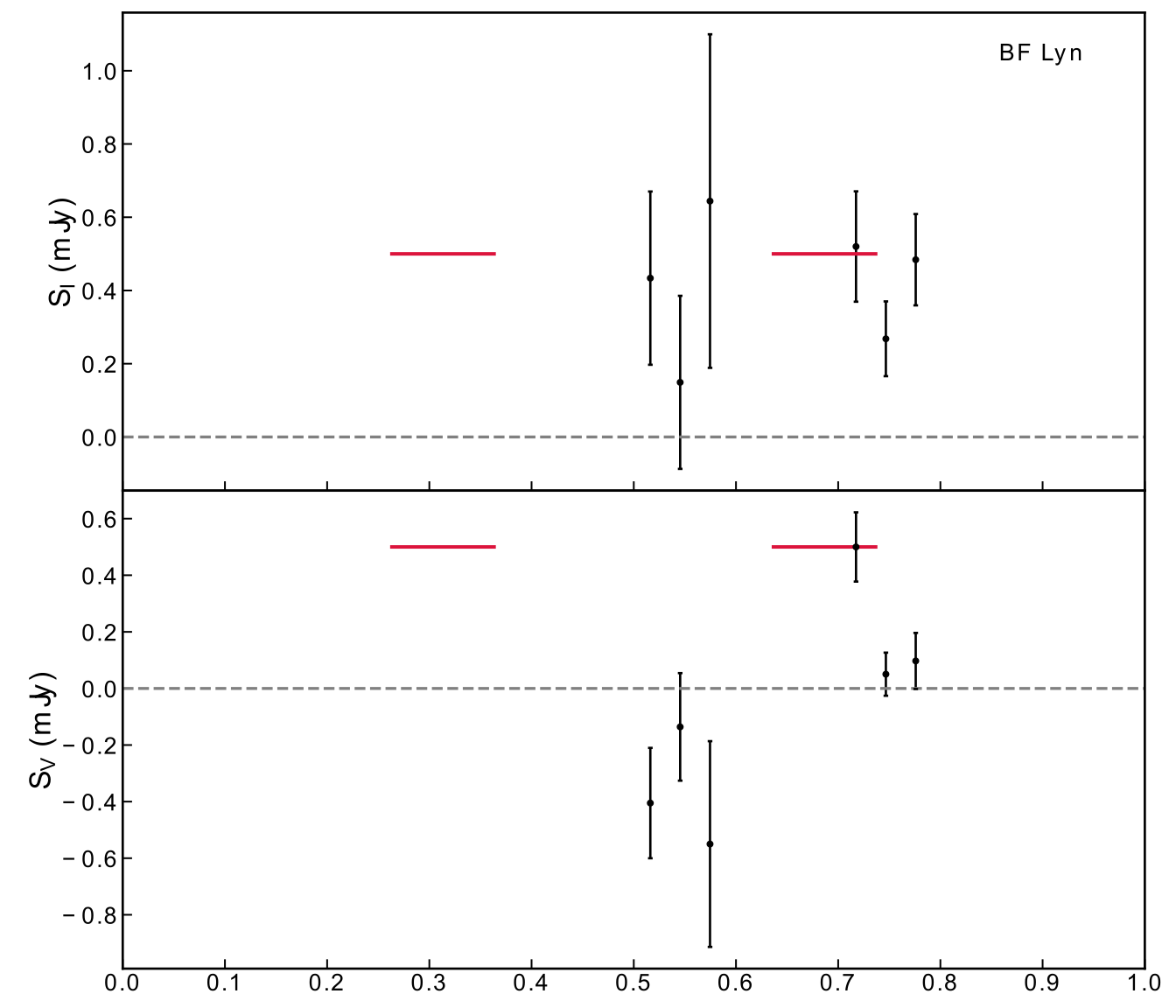} &
\includegraphics[scale=0.24]{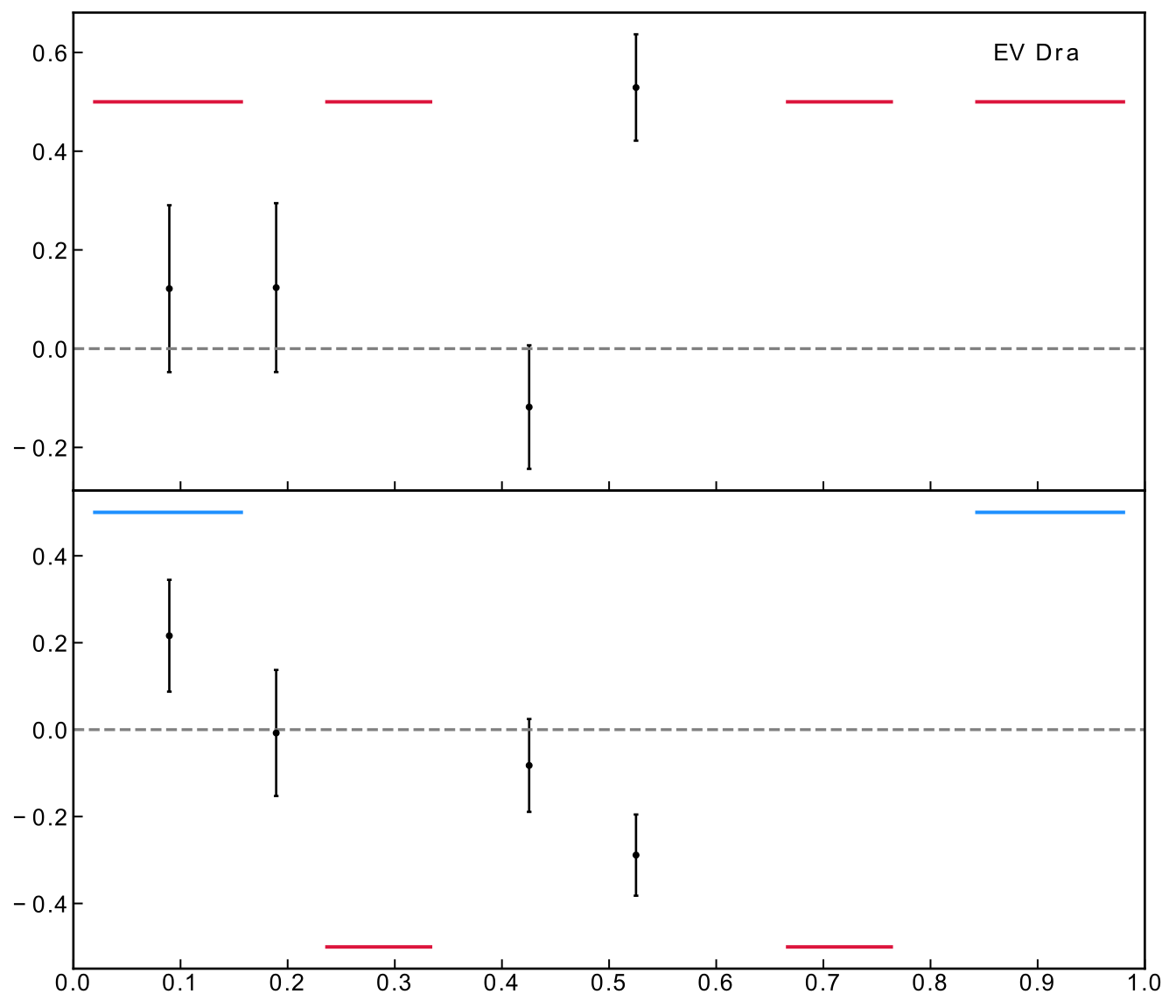} \\
\includegraphics[scale=0.24]{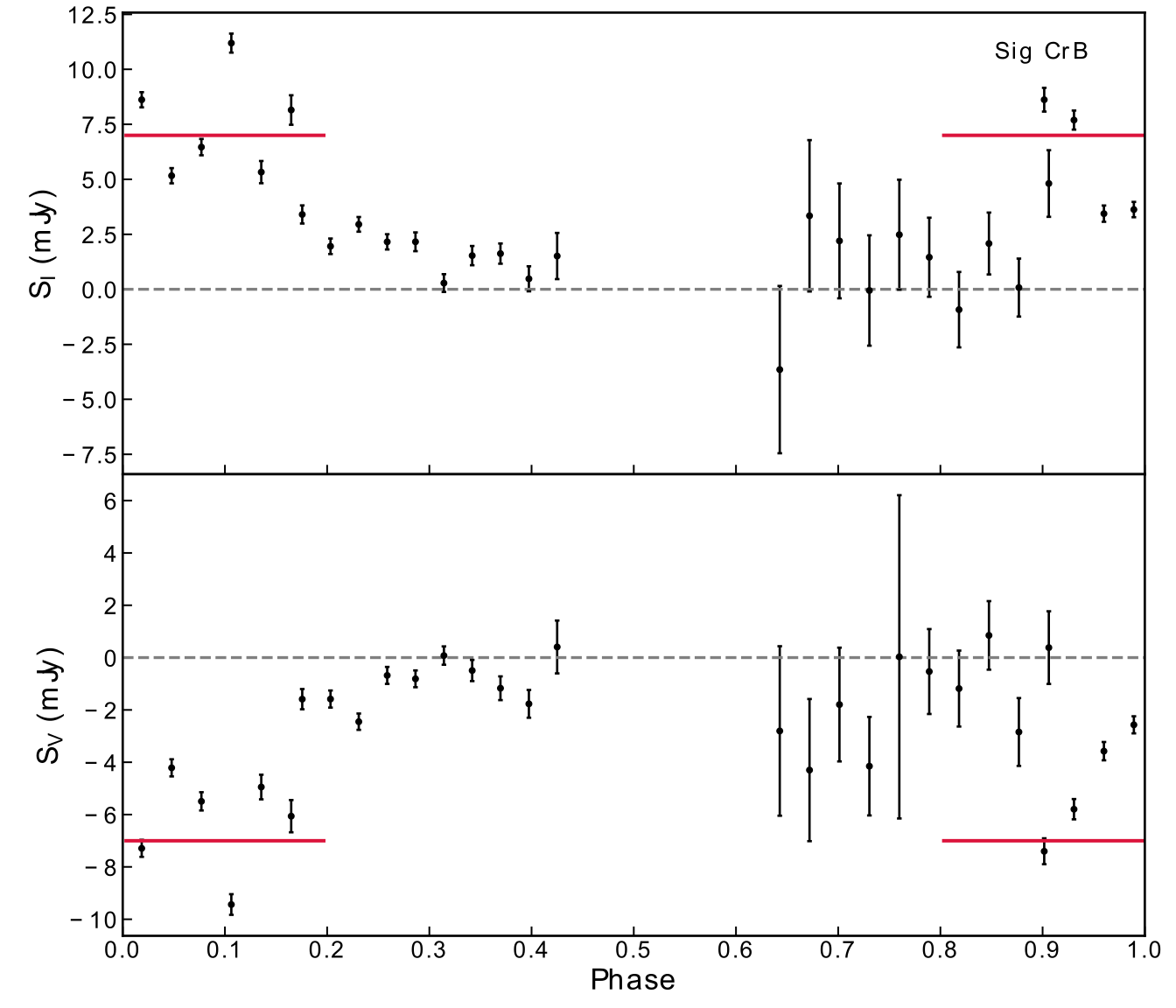} &
\includegraphics[scale=0.24]{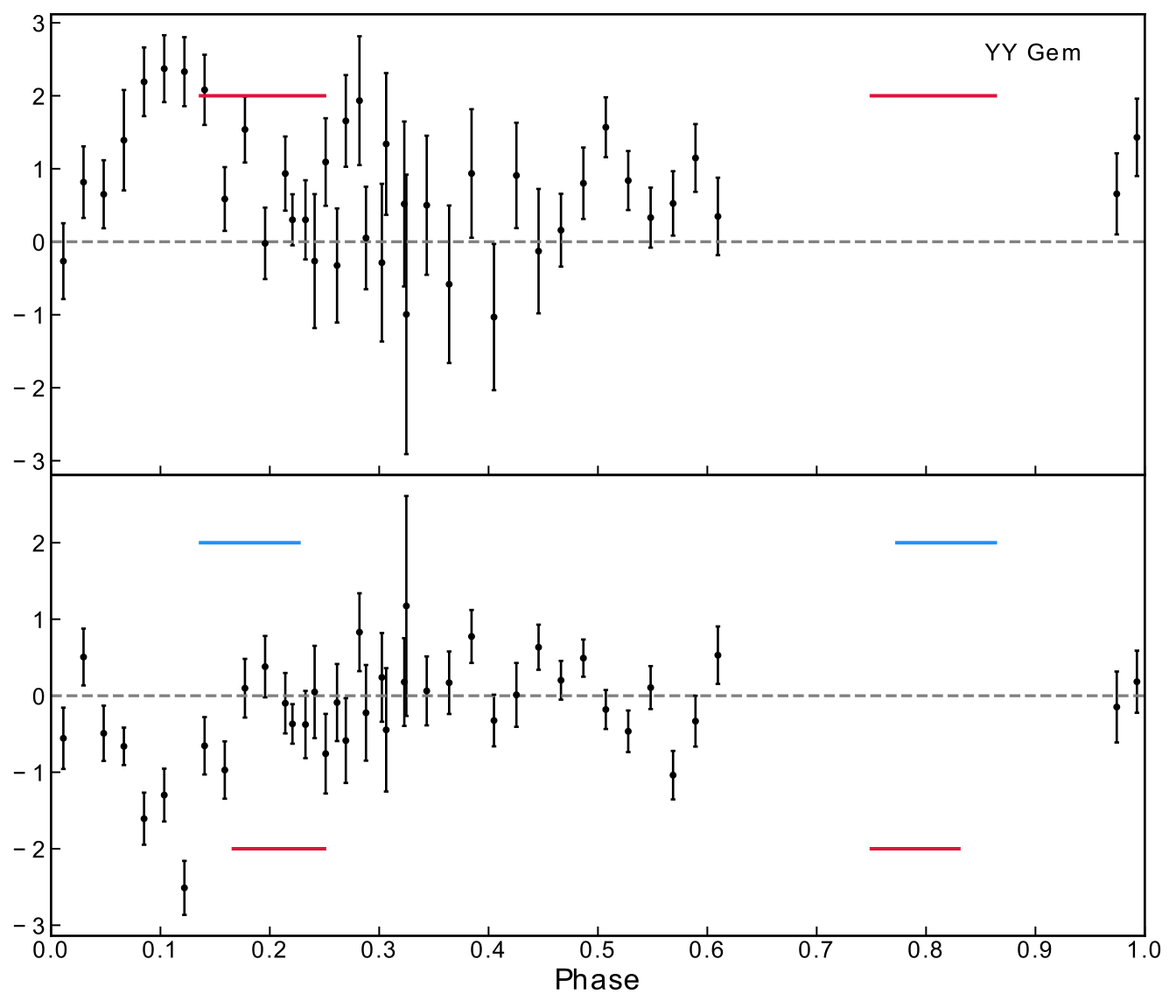}\\
\end{array}$
 \caption{Orbital phase of the low-frequency total intensity ($S_{I}$) and circularly-polarised ($S_{V}$) emission for BF\,Lyn (top left), EV\,Dra (top right), Sig\,CrB (bottom left) and YY\,Gem (bottom right). The orbital phase for which our model predicts we should observe emission is shown as red-lines at arbitrary flux densities. For Stokes V we use red and blue to distinguish between the two polarisations. Sig\,CrB was observed in three independent 8\,hr epochs, while all other stellar systems were only observed in two independent epochs. The prediction for Sig\,CrB has been shifted 0.5 in phase, which corresponds with emission originating on the secondary, to show the good correspondence with the data. The polarity of the Stokes V prediction is arbitrary because we do not know which magnetic pole is pointed towards us. Therefore the prediction for the stars where we see emission from both poles (YY\,Gem and EV\,Dra) can be flipped in sign.}
\label{fig:phasedig}
\end{center}
\end{figure*}

\begin{figure*}
    \centering
    \includegraphics[width=\linewidth]{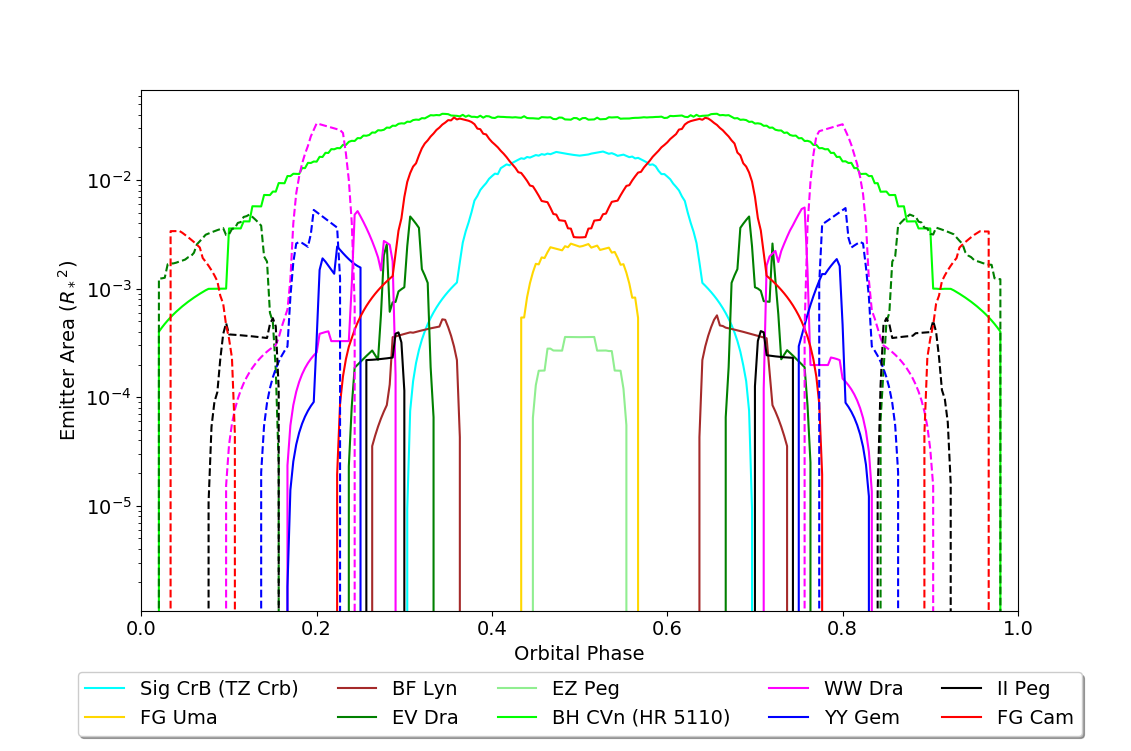}
    \caption{The predicted visible emitter area as a function of orbital phase for our targets that have measured inclinations. We have assumed a theoretical model for the emission geometry given by \citep{MelroseandDulk1982} with electron speed $\beta = 0.2$ as an estimate. The `step'-like variations between adjacent points are gridding artefacts. We have assumed that the observer's line of sight always hits the northern magnetic hemisphere of the emitting star. Solid lines are for electrons travelling from the north magnetic pole to the south magnetic pole, dashed lines are for electrons travelling vice versa. Since we do not know if we are observing the northern or southern magnetic hemisphere, these definitions are interchangeable. BQ CVn, DM UMa, DG CVn and BD+334462 have been left out because we lacked the necessary parameters to make an accurate prediction (namely the semi-major axis for the first three star systems, and the inclination for the last one).}
    \label{Emitter_Area_vs_Orbital_Phase}
\end{figure*}

The first system in our subset is Sigma$^2$ Corona Borealis (TZ\,CrB, HD\,146361), a F9V+G0V type star. It has an orbital period of $1.13979045 \pm 0.00000008$ days. \citet{Sig_Crb_Ephemeris} find that conjunction, with the secondary (G0V) closest to us, happens at JD = $2450127.9054 \pm 0.0004$. From this we can determine the orbital phase during the LOFAR epochs. They can be seen in Figure \ref{fig:phasedig}. Our prediction is shown in Figure \ref{Emitter_Area_vs_Orbital_Phase}. We see that our prediction aligns nicely with the observations if we shift the prediction 0.5 in phase. The shifted prediction is shown as the red line in Figure \ref{fig:phasedig}. A 0.5 phase shift is equivalent to saying that the emission originates on the 'companion' in our model, i.e. the secondary (G0V). This alignment is consistent with the magnetospheric acceleration model.

\subsubsection{BF\,Lyn}
The second star we studied is BF\,Lynx (HD\,80715), a K2V+dK type star. It has an orbital period of $3.804067 \pm 0.000008$ days. \citet{BF_Lyn_Ephemeris} find that conjunction happens at JD = $2447456.2383 \pm 0.0025$. The only credible detection here overlaps with our prediction, but the phase-binned data lacks a sufficient signal-to-noise ratio to draw a secure conclusion.

\subsubsection{EV\,Dra}
The third star in our subset is EV\,Draconis (HD\,144110), a G5V+K0V type star. It has an orbital period of 1.67140121 +- 0.00000065 days. \citep{EV_Dra_Spectral_Type} find that maximum radial velocity for the primary (G5V) happens at JD = $2452182.0226 \pm 0.0006$. This corresponds to an orbital phase of $0.75$. EV\,Dra's radio detection is offset by a phase of $0.25$ with respect to the predictions, which is inconsistent with the electrodynamic interaction model but consistent with the chromospheric acceleration scenario.

\subsubsection{YY\,Gem}
The fourth and final star in our subset is YY\,Geminorum (Castor\,C, HD\,60179C), a M1Ve+M1Ve type star. It has an orbital period of 0.81428290 ± 0.00000005 days. \citet{YY_Gem_Ephemeris} find that conjunction happens at BJD = $2456024.67475 \pm 0.00006$. The stars have the same spectral type, so there is no point in speaking of primary and secondary. This is the most puzzling case. The predicted epoch of visibility is offset in phase by $\approx 0.1$ of the rotation period, which neither supports the magnetospheric nor the chromospheric acceleration mechanism.

\section{Summary \& outlook}
We have uncovered a population of highly circularly polarised radio-emitting RS\,CVn stars. We argue based on the low frequency of detection ($144$\,MHz), high circular polarisation fraction ($\gtrsim 70\%$) and high implied brightness temperatures ($\gtrsim 10^{13}\,{\rm K}$) that we are seeing cyclotron maser emission from these systems. All previous radio detections of RS\,CVn systems were carried out at gigahertz frequencies, and with the notable exception of \citet{HR_1099}, interpreted as incoherent gyrosynchrotron emission. We argue that the surface strength of the large scale magnetic fields of RS\,CVn systems is likely $\sim 100\,{\rm G}$ which allows us to systematically probe their cyclotron emission for the first time due to the low frequency of our observations.

We find that the commonly-invoked loss cone maser can only account for the inferred brightness temperature for unreasonably large source sizes at any given frequency (of the order $\sim R_\ast^2$) for a significant fraction of our detections. We therefore suggest that the radio emission is being driven by the horseshoe or shell-type instability that also power radio emission in the Earth's magnetosphere. These types of masers require field-aligned potentials that accelerate charges. It is unclear if such a potential difference could be maintained in the dense corona of a star (plasma flows tend to `short' such voltage drops) and further theoretical work is necessary to gauge the feasibility of such a maser in stellar coronae.

We propose two plausible mechanisms that accelerate the cyclotron emitting electrons: enhanced chromospheric flaring due to tidal forces and an electrodynamical interaction between the two stars (similar to the Jupiter-Io interaction). We propose that quasi-contemporaneous radio monitoring and star-spot photometry could be used to conclusively establish the mechanism of acceleration. To this end, we have developed a code to predict the visibility of the emission and the projected source area for given stellar parameters (details in the Appendix). The limited radio data in hand do not point to a clear conclusion as to the mechanism of acceleration. 

We end up noting that the orbital ephemeris of some of our detections were determined using observations that are decades old. The error on the orbital period is often too large to predict the orbital phase during the newer LOFAR exposures sufficiently accurately to be able to constrain the mechanism of acceleration. In addition, some of the systems in our sample have unknown orbital inclination--- a necessary parameter to predict radio visibility. We therefore urge photometric observations of our targets by advanced amateur astronomers \citep[see e.g. ][]{gary} and spectroscopic observations by 1-metre class telescopes to accurately constrain the ephemeris, inclination and the longitude of active regions.

\bibliographystyle{aa}
\bibliography{ref}
\begin{acknowledgements}
JRC thanks the Nederlandse Organisatie voor Wetenschappelijk Onderzoek (NWO) for support via the Talent Programme Veni grant.
This paper is based on data obtained with the International LOFAR Telescope as part of the LoTSS survey. LOFAR is the Low Frequency Array designed and constructed by ASTRON. It has observing, data processing, and data storage facilities in several countries, that are owned by various parties (each with their own funding sources), and that are collectively operated by the ILT foundation under a joint scientific policy. The ILT resources have benefitted from the following recent major funding sources: CNRS-INSU, Observatoire de Paris and Universit\'{e} d'Orl\'{e}ans, France; BMBF, MIWF-NRW, MPG, Germany; Science Foundation Ireland (SFI), Department of Business, Enterprise and Innovation (DBEI), Ireland; NWO, The Netherlands; The Science and Technology Facilities Council, UK. This research made use of the Dutch national e-infrastructure with support of the SURF Cooperative (e-infra 180169) and the LOFAR e-infra group. The J\"{u}lich LOFAR Long Term Archive and the German LOFAR network are both coordinated and operated by the J\"{u}lich Supercomputing Centre (JSC), and computing resources on the supercomputer JUWELS at JSC were provided by the Gauss Centre for Supercomputing e.V. (grant CHTB00) through the John von Neumann Institute for Computing (NIC). AD acknowledges support by the BMBF Verbundforschung under the grant 05A20STA. This research made use of the University of Hertfordshire high-performance computing facility and the LOFAR-UK computing facility located at the University of Hertfordshire and supported by STFC [ST/P000096/1], and of the Italian LOFAR IT computing infrastructure supported and operated by INAF, and by the Physics Department of Turin university (under an agreement with Consorzio Interuniversitario per la Fisica Spaziale) at the C3S Supercomputing Centre, Italy. {\em Software used: } \texttt{python}, \texttt{numpy}, \texttt{scipy}, \texttt{astropy}, \texttt{matplotlib}, \texttt{vector3d}. {\em Telescopes used: } LOFAR
\end{acknowledgements}


\begin{appendix}
\section{Cyclotron maser geometry}
Here we describe our modelling of the geometry of the loss cone maser. 
We begin with the following assumptions:
\begin{enumerate}
    \item The rotational and orbital axis of the binaries are aligned. This assumption is justified for synchronous binaries ($P_{rot} = P_{orb}$) but not for asynchronous ones ($P_{rot}\neq P_{orb}$) \citep{RandOAxisAlign}. As can be seen in Table \ref{Table_Orbital_Parameters},  all our binaries except FG Cam (and potentially BH CVn and DG CVn as the rotational period and orbital period respectively are unknown) are synchronous.
    \item The stellar magnetic field resembles a dipole and its magnetic axis aligns with the rotational  axis \citep{RS_CVn_Magnetic_Field_Topology}. The surface magnetic field at the equator is taken to be 100\,G which is a representative value for RS\,CVn systems \citep{RS_CVn_typical_field_strength}.
    \item The orbit is circular. As can be seen in Table \ref{Table_Orbital_Parameters}, the eccentricity is 0 or unknown for all binaries except BH CVn. 
    \item The wind from the companion does not significantly distort the magnetic topology of the primary star near the emission site. This may not be true in a subset of close contact binaries but we retain this assumption for simplicity.
\end{enumerate}

We model the magnetic field and emission geometry on a grid of points in the magnetosphere of the emitter. 
We refer to the region where the magnetic field has the right value to generate emission at a given frequency range (and the field lines intercept the binary companion in the case of binary interaction) as the $\emph{source site}$. A sub-set of this region from where the beamed emission reaches the Earth is referred to as the $\emph{emitter site}$. 

We adopted the beaming geometry of \citet{MelroseandDulk1982} where the emission is directed into a hollow cone with opening angle $\theta_0$ and thickness $\Delta\theta_0$ where $\cos\theta_0 \approx \Delta\theta_0\approx \beta$. In reality, the precise beaming geometry of the emergent radiation depends on the nature of the instability and refraction near the emitter which are not directly accessible, so we take these as representative values. Similarly, we assume that the emitting electrons have an energy of about 10\,keV ($\beta\approx 0.2$) as a representative value. With these values, any grid point in the {\rm source site} where the magnetic field vector makes an angle between $\theta_0-\Delta\theta_0/2$ and $\theta_0+\Delta\theta_0/2$ with the vector pointing towards the observer will fall within the emitter site.

We also adopted the maser bandwidth from \citet{MelroseandDulk1982}: $\Delta \nu/\nu \approx 0.5\beta^2\cos^2\alpha_0$, where $\alpha_0$ is the loss-cone angle. We considered $\alpha_0=0.5$ as a representative value, which along with $\beta=0.2$ gives $\Delta\nu\approx 2\,{\rm MHz}$ in the LOFAR band. Because we compared the inferred brightness temperature with the theoretical value for a single maser site, we restricted the source site to a region whose cyclotron frequency is within $\Delta \nu$ of the central observational frequency of $144$\,MHz. 
Finally, we computed the area of the emitter for brightness temperature calculation by projecting the three-dimensional emitter site on to a plane that is normal to the line of sight to Earth. 

\begin{figure}
    \centering
    \includegraphics[width=\linewidth]{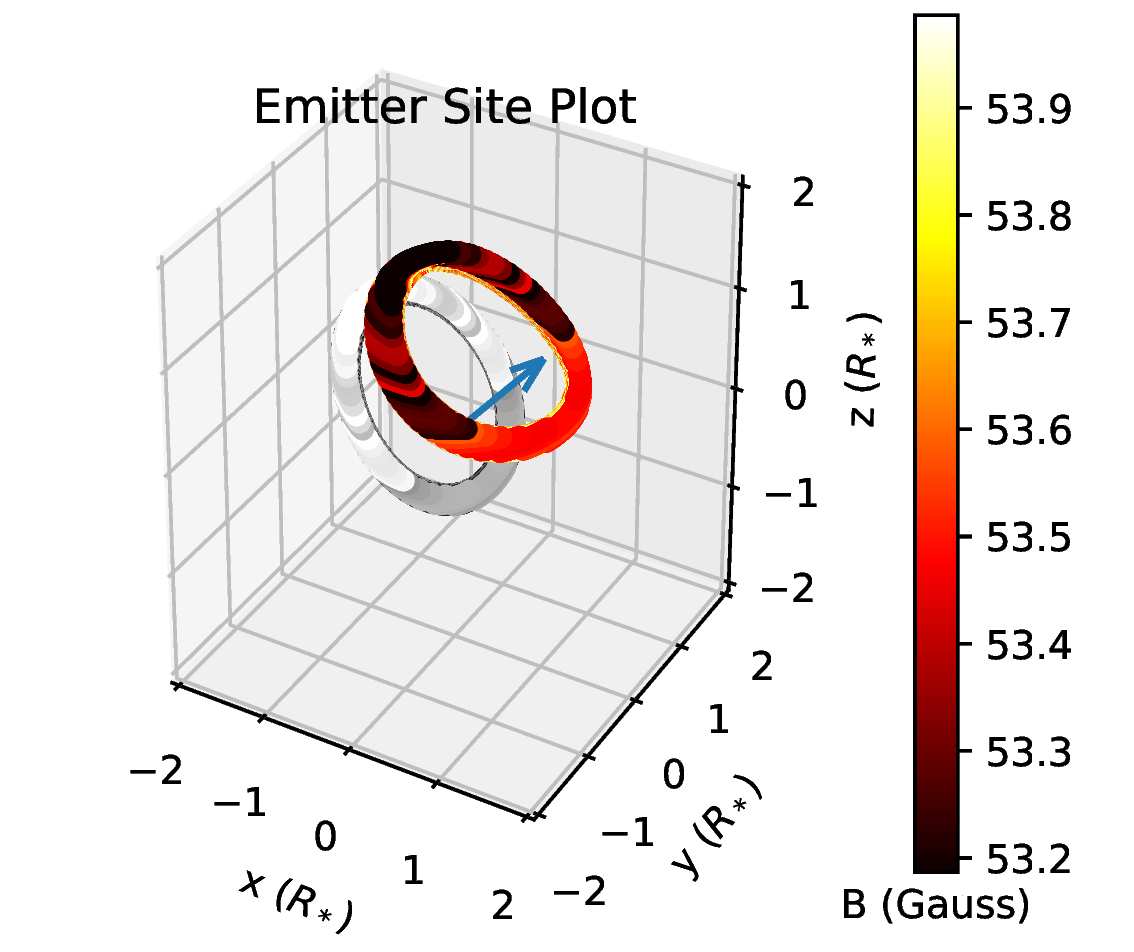}
    \caption{The emitter site in colour with beaming taken into account. The blue arrow points towards the observer. $\beta$ is 0.2 and the observing angle is 45 degrees in this example. The transverse emitter site is depicted in black and white. The star is located at the origin.}
    \label{Emitter_Site}
\end{figure}

\begin{figure}
    \centering
    \includegraphics[width=\linewidth]{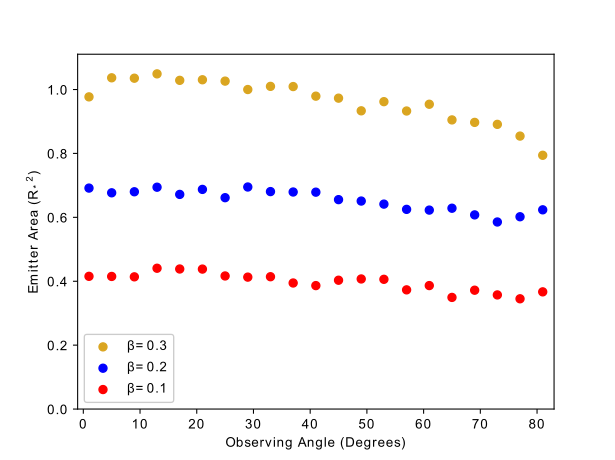}
    \caption{The transverse emitter site area as a function of $\beta$ and observer angle. $\beta$ has a much larger effect than the observer angle. This is due to the fact that the thickness of the emission cone gets wider with $\beta$.}
    \label{Emitter_Site_vs_observing_angle_and_beta}
\end{figure}

Fig. \ref{Emitter_Site} shows the emitter site for an example run where we have assumed that the entire magnetosphere of the star is live with cyclotron emitting electrons. Fig. \ref{Emitter_Site_vs_observing_angle_and_beta} shows the variation of the projected emitter size with $\beta$ and the observer angle, defined as the inclination of the system. Our numerical calculation for the size of the emitter agrees with the approximate expression $A=\pi\beta R_\ast^2$ used in the main text.

To model the binary interaction scenario we additionally restrict the {\em source site} to the flux tube connecting the radio emitting star with its companion.
To determine whether a point is on a magnetic field line passing through the companion, we use the equation for magnetic streamlines of a dipole: $l=r\sin^2\theta$, where $\theta$ is the polar angle, $l$ is the distance at which the field line crosses the equator and $r$ is the radial distance. If $R_2$ is the radius of the companion, then only field lines that have equatorial intercepts in the range 
\begin{equation}
     l\in[a-R_2/2,\,a+R_2/2],
     \label{Equation_Field_Lines_Through_Binary}
\end{equation}
can be live. Similarly, in the azimuthal direction, only field lines within 
\begin{equation}
    -\tan^{-1}\left(\frac{R_2}{a}\right) \leq \phi \leq \tan^{-1}\left(\frac{R_2}{a}\right),
    \label{Equation_Phi_space}
\end{equation}
where $\phi$ is the azimuthal angle, can be live. Equations \ref{Equation_Field_Lines_Through_Binary} and \ref{Equation_Phi_space} completely determine the live flux tube in case of binary interaction. Figure \ref{Binary_Interaction_Gifs_Magnetic_Field_Lines} shows the field lines of the central star passing through the binary companion. Figure \ref{Binary_Interaction_Gifs_Source_And_Emitter_Sites} shows the source and emitter sites. The grid was set to a step size of 0.003 in units of $R_*$ for r, and 0.003 radians for $\theta$ and $\phi$.
\\
\begin{figure*}
    \begin{center}
        \subfigure{\animategraphics[autoplay,loop,width=0.49\linewidth,height=0.4\linewidth]{8}{Sig_CrB_TZ_CrB_Magnetic_Field_Lines/"Magnetic_Field_Lines_Through_Binary_Sig_CrB_TZ_Crb_BR_0.96_BD_5.21_28_degrees_0point2c_frame"}{0}{31}}
        \subfigure{\animategraphics[autoplay,loop,width=0.49\linewidth,height=0.4\linewidth]{8}{Sig_CrB_TZ_CrB_Magnetic_Field_Lines_Zoomed_In/"Magnetic_Field_Lines_Through_Binary_Zoomed_In_On_Central_Star_Sig_CrB_TZ_Crb_BR_0.96_BD_5.21_28_degrees_0point2c_frame"}{0}{31}}
    \end{center}
    \caption{Simulations showing the magnetic field lines from the central star passing through the binary. In the left image we have zoomed out so the binary is constantly visible. The image on the right is zoomed in on the central star because the region we are interested in, with $\bf{|B|}\approx $ 50 Gauss, is located there. Normally as described in the text the magnetic field strength criterion is applied first, but for illustrative purposes we show that in the next figure. The blue arrow depicts the direction towards the observer, its angle corresponds with the inclination of the system. When viewed in Adobe Acrobat Reader these images will appear as gifs, showcasing the fact that the celestial bodies orbit around each other.}
    \label{Binary_Interaction_Gifs_Magnetic_Field_Lines}
\end{figure*}

We have also incorporated the orbital motion of the binary system. This brings along the time variability that we see in our data. As can be seen in the right gif in Figure \ref{Binary_Interaction_Gifs_Source_And_Emitter_Sites} (animation may be viewed using Adobe Reader), the emitter site is only visible at certain orbital phases. At other phases, when it is not visible, we place a single point with magnetic field strength 0 at the origin. This makes it immediately clear from the colour bar there is no emitter site at that particular orbital phase.

\begin{figure*}
    \begin{center}
        \subfigure{\animategraphics[autoplay,loop,width=0.49\linewidth,height=0.4\linewidth]{8}{Sig_CrB_TZ_CrB_Source_Site/"Source_Site_Binary_Interaction_NtoS_Sig_CrB_TZ_Crb_BR_0.96_BD_5.21_28_degrees_0point2c_frame"}{0}{31}}
        \subfigure{\animategraphics[autoplay,loop,width=0.49\linewidth,height=0.4\linewidth]{8}{Sig_CrB_TZ_CrB_Emitter_Site/"Emitter_Site_Binary_Interaction_NtoS_Sig_CrB_TZ_Crb_BR_0.96_BD_5.21_28_degrees_0point2c_frame"}{0}{31}}
    \end{center}
    \caption{Simulations showing the source and emitter sites. In the image on the left we have applied the criterion that the magnetic field strength converted to frequency must be inside the bandwidth of a single maser site. In the image on the right we have taken into account on top of the bandwidth criterion that the radiation is beamed in a hollow cone with a solid rim. Therefore the emitter is not always visible. The blue arrow depicts the direction towards the observer, its angle corresponds with the inclination of the system. When viewed in Adobe Acrobat Reader these images will appear as gifs, showcasing the fact that the celestial bodies orbit around each other. Only in certain frames the emitter site will be visible. If it is not visible, a single point with magnetic field strength 0 is placed at the origin so it is clear from the colour bar that the emitter site is not visible at this phase.}
    \label{Binary_Interaction_Gifs_Source_And_Emitter_Sites}
\end{figure*}

\begin{table}[]
    \centering
    \begin{tabular}{|l|l|}
    \hline
    Reference Number & Reference \\ \hline
        \setcitestyle{numbers} \citep{WW_Dra_Rot_Per} & \setcitestyle{authoryear} \citep{WW_Dra_Rot_Per} \\ \hline 
        \setcitestyle{numbers} \citep{II_Peg_Spectral_Type} & \setcitestyle{authoryear} \citep{II_Peg_Spectral_Type} \\ \hline 
        \setcitestyle{numbers} \citep{Gaia_DR2} & \setcitestyle{authoryear} \citep{Gaia_DR2} \\ \hline
        \setcitestyle{numbers} \citep{Gaia_DR1} & \setcitestyle{authoryear} \citep{Gaia_DR1} \\ \hline 
        \setcitestyle{numbers} \citep{BH_CVn_eccentricity} & \setcitestyle{authoryear} \citep{BH_CVn_eccentricity} \\ \hline 
        \setcitestyle{numbers} \citep{DM_UMa_Mass_Hot_Component} & \setcitestyle{authoryear}  \citep{DM_UMa_Mass_Hot_Component} \\ \hline 
        \setcitestyle{numbers} \citep{RS_CVn_typical_field_strength} & \setcitestyle{authoryear} \citep{RS_CVn_typical_field_strength} \\ \hline 
        \setcitestyle{numbers} \citep{BH_CVn_Spectral_Type} & \setcitestyle{authoryear} \citep{BH_CVn_Spectral_Type} \\ \hline 
        \setcitestyle{numbers} \citep{BQ_CVn_Radius_Cold_Component} & \setcitestyle{authoryear} \citep{BQ_CVn_Radius_Cold_Component} \\ \hline 
        \setcitestyle{numbers} \citep{FG_UMa_Radius_Cold_Component} & \setcitestyle{authoryear} \citep{FG_UMa_Radius_Cold_Component} \\ \hline 
        \setcitestyle{numbers} \citep{EV_Dra_Spectral_Type} & \setcitestyle{authoryear} \citep{EV_Dra_Spectral_Type} \\ \hline 
        \setcitestyle{numbers} \citep{WW_Dra_Radii} & \setcitestyle{authoryear} \citep{WW_Dra_Radii} \\ \hline 
        \setcitestyle{numbers} \citep{WW_Dra_Spectral_Type} & \setcitestyle{authoryear} \citep{WW_Dra_Spectral_Type} \\ \hline 
        \setcitestyle{numbers} \citep{BD+334462_Spectral_Type} & \setcitestyle{authoryear} \citep{BD+334462_Spectral_Type} \\ \hline 
        \setcitestyle{numbers} \citep{YY_Gem_Ephemeris} & \setcitestyle{authoryear} \citep{YY_Gem_Ephemeris} \\ \hline
        \setcitestyle{numbers} \citep{RandOAxisAlign} & \setcitestyle{authoryear} \citep{RandOAxisAlign} \\ \hline 
        \setcitestyle{numbers} \citep{EZ_Peg_Orb_Per} & \setcitestyle{authoryear} \citep{EZ_Peg_Orb_Per} \\ \hline 
        \setcitestyle{numbers} \citep{BQ_CVn_Spectral_Type} & \setcitestyle{authoryear} \citep{BQ_CVn_Spectral_Type} \\ \hline 
        \setcitestyle{numbers} \citep{FG_Cam_Spectral_Type} & \setcitestyle{authoryear} \citep{FG_Cam_Spectral_Type} \\ \hline 
        \setcitestyle{numbers} \citep{DM_UMa_Radius_Hot_Component} & \setcitestyle{authoryear} \citep{DM_UMa_Radius_Hot_Component} \\ \hline 
        \setcitestyle{numbers} \citep{BF_Lyn_Ephemeris} & \setcitestyle{authoryear} \citep{BF_Lyn_Ephemeris} \\ \hline
        \setcitestyle{numbers} \citep{DM_UMa_Rot_Per} & \setcitestyle{authoryear} \citep{DM_UMa_Rot_Per} \\ \hline 
        \setcitestyle{numbers} \citep{BQ_CVn_Rot_Per} & \setcitestyle{authoryear} \citep{BQ_CVn_Rot_Per} \\ \hline 
        \setcitestyle{numbers} \citep{BH_CVn_Masses} & \setcitestyle{authoryear} \citep{BH_CVn_Masses} \\ \hline 
        \setcitestyle{numbers} \citep{FG_UMa_Spectral_Type} & \setcitestyle{authoryear} \citep{FG_UMa_Spectral_Type} \\ \hline 
        \setcitestyle{numbers} \citep{BF_Lyn_Radii} & \setcitestyle{authoryear} \citep{BF_Lyn_Radii} \\ \hline 
        \setcitestyle{numbers} \citep{BF_Lyn_Spectral_Type} & \setcitestyle{authoryear} \citep{BF_Lyn_Spectral_Type} \\ \hline 
        \setcitestyle{numbers} \citep{EZ_Peg_Spectral_Type} & \setcitestyle{authoryear} \citep{EZ_Peg_Spectral_Type} \\ \hline 
        \setcitestyle{numbers} \citep{DM_UMa_Spectral_Type} & \setcitestyle{authoryear} \citep{DM_UMa_Spectral_Type} \\ \hline 
        \setcitestyle{numbers} \citep{DG_CVn_Spectral_Type} & \setcitestyle{authoryear} \citep{DG_CVn_Spectral_Type} \\ \hline 
        \setcitestyle{numbers} \citep{FG_Cam_Rot_Per} & \setcitestyle{authoryear} \citep{FG_Cam_Rot_Per} \\ \hline 
        \setcitestyle{numbers} \citep{WW_Dra_Masses} & \setcitestyle{authoryear} \citep{WW_Dra_Masses} \\ \hline 
        \setcitestyle{numbers} \citep{YY_Gem_Spectral_Type} & \setcitestyle{authoryear} \citep{YY_Gem_Spectral_Type} \\ \hline 
        \setcitestyle{numbers} \citep{Sig_Crb_Ephemeris} & \setcitestyle{authoryear} \citep{Sig_Crb_Ephemeris} \\ \hline
        \setcitestyle{numbers} \citep{BF_Lyn_Rot_Per} & \setcitestyle{authoryear} \citep{BF_Lyn_Rot_Per} \\ \hline 
        \setcitestyle{numbers} \citep{Sig_CrB_Spectral_Type} & \setcitestyle{authoryear} \citep{Sig_CrB_Spectral_Type} \\ \hline
        \setcitestyle{numbers} \citep{II_Peg_Rot_Per} & \setcitestyle{authoryear} \citep{II_Peg_Rot_Per} \\ \hline 
        \setcitestyle{numbers} \citep{BF_Lyn_Masses} & \setcitestyle{authoryear} \citep{BF_Lyn_Masses} \\ \hline 
        \setcitestyle{numbers} \citep{YY_Gem_Masses} & \setcitestyle{authoryear} \citep{YY_Gem_Masses} \\ \hline 
        \setcitestyle{numbers} \citep{Hipparcos} & \setcitestyle{authoryear} \citep{Hipparcos} \\ \hline 
    \end{tabular}
    \caption{Reference Table for the references being used in the stellar and orbital parameters Tables (Tables \ref{Table_Stellar_Parameters} and \ref{Table_Orbital_Parameters}).}
    \label{Reference_table}
\end{table}

\end{appendix}
\end{document}